\documentclass[aps,prl,twocolumn,superscriptaddress,showpacs,floatfix]{revtex4-2}
\usepackage[utf8]{inputenc}
\usepackage[T1]{fontenc}
\usepackage{float} 
\usepackage{color}  
\usepackage{graphicx}
\usepackage{amsmath}
\usepackage{amssymb}
\usepackage{bbm}
\usepackage{indentfirst}
\usepackage{dcolumn}
\usepackage{soul}
\usepackage{mathrsfs}
\usepackage{xcolor}

\usepackage{relsize}
\usepackage{amsmath}
\usepackage{graphicx}
\usepackage{float}
\usepackage{color}
\makeatletter

\usepackage{xcolor}
\usepackage{soul}
\usepackage{soulpos}

\usepackage{dcolumn}
\usepackage{bm}
\usepackage{subfigure}
\usepackage{helvet}
 \usepackage{color}
\graphicspath{{figures/}}

\usepackage{tcolorbox}

\usepackage[colorlinks=true,linkcolor=blue,citecolor=blue,urlcolor=blue]{hyperref}

\begin{document}

\preprint{APS/123-QED}

\title{Electron-phonon mediated spin-flip as driving mechanism for ultrafast magnetization dynamics in 3$d$ ferromagnets}

\author{Theodor Griepe}
\affiliation{Dahlem Center for Complex Quantum Systems and Fachbereich Physik, Freie Universit\"{a}t Berlin, 14195 Berlin, Germany}
\affiliation{Instituto de Ciencia de Materiales de Madrid, CSIC, Cantoblanco, 28049 Madrid, Spain\looseness=-1}
\author{Unai Atxitia}%
\email{u.atxitia@csic.es}
\affiliation{Dahlem Center for Complex Quantum Systems and Fachbereich Physik, Freie Universit\"{a}t Berlin, 14195 Berlin, Germany}
\affiliation{Instituto de Ciencia de Materiales de Madrid, CSIC, Cantoblanco, 28049 Madrid, Spain\looseness=-1}

\date{\today}

\begin{abstract}
Despite intense experimental effort, theoretical proposals and modeling approaches, a lack of  consensus exists about the intrinsic mechanisms driving ultrafast magnetization dynamics in 3$d$ ferromagnets.
In this work, we find evidence of electron-phonon mediated spin-flip as the driving mechanism for the ultrafast magnetization dynamics in all three 3$d$ ferromagnets; nickel, iron and cobalt. 
We use a microscopic three temperature model with  parameters calculated from first-principles, which has been validated by direct comparison to the electron and lattice dynamics extracted from previous experiments.
By direct comparison to the experimentally measured magnetization dynamics for different laser fluence, we determine the spin-flip probability of each material. 
In contrast to previous findings but in agreement to ab-initio predictions, we find that relatively small values of the spin-flip probability enable ultrafast demagnetization in all three 3$d$ ferromagnets.
\end{abstract}

\maketitle

The discovery of femtosecond spin dynamics in the 3$d$ ferromagnet nickel opened the door for magnetic non-volatile data processing on ultrafast time scales\cite{Beaurepaire1996}. Since then, ultrafast demagnetization has been demonstrated in other transition metals \cite{Carpene2008, Krauss2009}, rare-earth metals \cite{Wietstruk2011,Frietsch2015,Frietsch2020,Windsor2022} and their alloys \cite{Radu2011}, semiconductors \cite{Matsubara2015} and  insulators \cite{Maehrlein2019}. 
The fundamental mechanisms behind laser-induced ultrafast magnetization dynamics, such as the nature of spin-flip scattering or the role of spin-polarized transport, are still object of debate \cite{Koopmans2005,Koopmans2010,Battiato2010,Carva2011,Schellekens2013,Dewhurst2021}. 
In the first works, Koopmans et al. \cite{Koopmans2005,Koopmans2010} proposed electron-phonon scattering mediated spin-flip as the main mechanism driving the ultrafast magnetization dynamics in 3$d$ ferromagnets. 
This idea was supported by fitting the so-called microscopic three temperature model (M3TM) to ultrafast demagnetization traces in nickel and cobalt \cite{Koopmans2010}.
Later on, it was argued that other mechanisms should significantly contribute to the demagnetization since the ab-initio estimated values for the spin-flip rates were too small to produce the observed demagnetization \cite{Carva2011,Carva2013}.
The  main drawback of the M3TM as originally used by Koopmans et al. \cite{Koopmans2010} is the large number of fitting parameters. 
This has been partially solved in subsequent works in nickel \cite{Roth2012,Schellekens2013} where the temperature dependence of the electron, lattice and spin specific heats at equilibrium were directly taken from experiments\cite{Meschter1981}. 
By doing so, the spin-flip rates necessary to fit the experimental results reduced and approached to ab-initio estimations.  
Besides electron-phonon spin-flip scattering, other alternative mechanisms have been put forward, such as Coulomb scattering of Elliot-Yafet type \cite{Krauss2009} and electron-magnon scattering \cite{Carpene2008,Tveten2015}. The latter aims to describe angular momentum conservation due to an increase of the orbital momenta upon creation of a magnon. However, experimental studies  suggest that angular momentum dissipation to orbital momenta cannot explain the magnetization quenching in iron and cobalt on this timescale and support the claim of the lattice to absorb excess angular momentum\cite{Stamm2007, Cinchetti2006,Dornes2019,Tauchert2022}.
Superdiffusive spin currents have been also proposed as dominant mechanism driving the ultrafast demagnetization, however this process is only efficient when magnets are interfaced by nonmagnetic metals \cite{Battiato2010,Eschenlohr2013}. Whereas in thin films on insulating substrates, superdiffusive transport plays no significant role in the demagnetization process of 3$d$ ferromagnets \cite{Schellekens2013}.
We note that recent ab-initio time-dependent density functional theory have  shown to be accurate in the early stages of the optically induced demagnetization process before lattice vibrations set in \cite{Shokeen2017}. However, experimental evidence suggests that angular momentum from the magnetic system is transferred to the lattice \cite{Tauchert2022, Dornes2019}.
These processes are not within reach of ab-initio methods, and, therefore, their investigation needs to rely on semi-phenomenological models.

Femtosecond-resolved experimental techniques that  measure the magnetization dynamics as well as the electronic and lattice dynamics are nowadays accessible. 
The electron energy distribution has been measured in nickel using time- and angle-resolved photo-emission spectroscopy (tr-ARPES)  \cite{Tengdin2018,You2018}. It has been demonstrated that only a few tens of femtoseconds after the laser hits the sample, the electron system  thermalizes into a Fermi-Dirac distribution with a well-defined electron temperature. The lattice energy dynamics has been measured using time-resolved electron diffraction in the three 3$d$ ferromagnets \cite{ZahnNi,ZahnCo_Fe,Tauchert2022}.
It has been demonstrated that the so-called two- temperature model (2TM) describes well the measured electron and lattice temperature dynamics for parameters calculated from first principles, and by including the energy exchange between the spin and electron systems within an atomistic spin model framework \cite{ZahnNi,ZahnCo_Fe}. 
However, the lack of a theoretical model describing the magnetization dynamics questions the full validity of the atomistic spin model.

In this work, we demonstrate that the M3TM is able to quantitatively describe the ultrafast dynamics in 3$d$ ferromagnets, including electron, lattice and spin degrees of freedom. To do so, we replicate the dynamics of the experimentally measured electron and lattice temperatures as well as magnetization for a range of laser fluences in  all three 3$d$ ferromagnetic transition metals iron, nickel and cobalt. Notably, we are able to reduce the number of fit parameters to only the spin-flip probability. We achieve so by using system parameters that have been calculated from ab-initio methods and experimentally validated. By contrast to previous works, we find that the spin-flip probabilities agree with to those calculated via ab-initio \cite{Carva2011,Carva2013}. We conclude that our work evidences electron-phonon spin-flips as intrinsic mechanism driving ultrafast magnetization dynamics in 3$d$ ferromagnets. 

\begin{figure}[t!]
    \centering
    \includegraphics[width=0.9\linewidth]{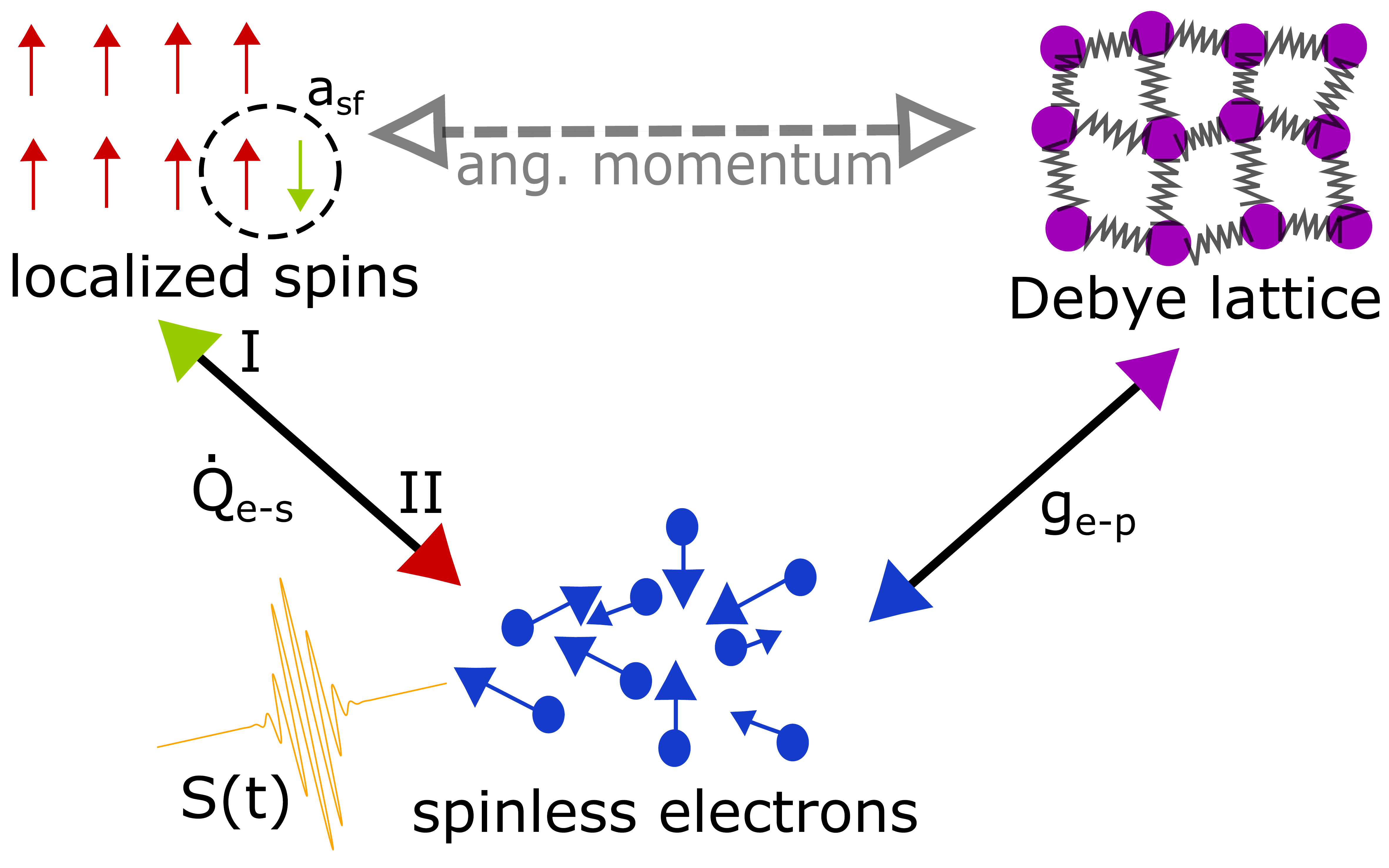}
    \caption{
    Energy absorbed by the spinless electrons $S(t)$ is distributed to the lattice ($g_{\rm{ep}}$) and localized spins ($\dot{Q}_{\rm{e-s}}$). Energy is transferred from electron to spin during demagnetization (I), and spin to electrons during remagnetization (II). The angular momentum from electron-phonon mediated spin-flips, with probability $a_{\rm{sf}}$, is implicitly exchanged with lattice.}
    \label{fig:model}
\end{figure}

Our model is based on an extension to the M3TM \cite{Koopmans2010,Beens2019}. The energy flow dynamics are described by the two-temperature model (see  Fig. \ref{fig:model}),
\begin{eqnarray}
{C_{\rm{e}}}   \frac{d T_{\rm{e}} }{dt} &=& {g_{\rm{ep}}} (T_{\rm{p}}-T_{\rm{e}}) + S(t)+ \dot{Q}_{\rm{e-s}} \label{eq:TTM1}\\
{C_{\rm{p}}}  \frac{d T_{\rm{p}}}{dt}  &=& -{g_{\rm{ep}}}(T_{\rm{p}}-T_{\rm{e}}).
\label{eq:TTM2}
\end{eqnarray}
When a metallic thin film is subjected to an optical laser pulse, only the electrons are excited by the photon electric field.  Initially, the absorbed energy is barely transferred to the lattice and consequently the electron system heats up. On the timescale of 100 fs, the electron temperature will rise far above the critical temperature, $T_{\rm{c}}$. 
The magnetic system responds to this fast temperature rise by reducing its magnetic order on similar time scales. 
The electron and lattice systems are assumed to be thermalized so that their energy can be described by a temperature, $T_{\rm{e}}$ for the electrons and $T_{\rm{p}}$ for the phonons.
In Eq. \eqref{eq:TTM1}, the absorbed laser pulse power is represented by a Gaussian function, $S(t)=S_0*G(\tau_p)$, where $\tau_p$ is the pulse duration. The electron heat capacity is $C_e$. The electron-phonon coupling allows for temperature equilibration of hot electrons and the lattice on the time scale determined by the ratio $g_{\rm{ep}}/C_{\rm{e}}$. Since $g_{\rm{ep}}$ accounts only for spin-conserving scattering events, in  Eq. \eqref{eq:TTM1} we include a term that accounts for the finite energy cost (gain) of a spin-flip and couple it to the electron dynamics: $\dot{Q}_{\rm{e-s}}= J_0 m \dot{m} / V_{\rm{at}}$, with $J_0/3=[S/(S+1)]  k_{\rm{B}}T_c$ in the MFA, $m$ is the reduced magnetization, $J_0$ is the exchange energy, $V_{\rm{at}}$ the atomic volume and effective spin $S$. In phase I ($dm/dt<0$) the energy cost of an electron-phonon mediated spin-flip of probability $a_{\rm{sf}}$ in minority direction is deducted from the electronic energy, while in phase II ($dm/dt>0$) the direction of energy flow is reverted (Fig. \ref{fig:model}).
In other works the energy flow between the spin and the electron systems has been taken into account by adding it to the electron specific heat, $C_{\rm{e}} \rightarrow C_{\rm{e}}+C_{\rm{s}}$, where $C_{\rm{s}}$ is the equilibrium spin specific heat \cite{ZahnNi, Tengdin2018}. However, recent works suggest that this energy flow needs to be calculated through the spin Hamiltonian  for non-equilibrium spin configurations \cite{ZahnNi, ZahnCo_Fe}. 
An important aspect of the two-temperature model in Eqs. \eqref{eq:TTM1} and \eqref{eq:TTM2} is the exact value of $C_{\rm{e}}$, $C_{\rm{p}}$, and $g_{\rm{ep}}$. In this work, we use the ab-initio calculated parameters, which were already used before in Ref. \cite{ZahnNi,ZahnCo_Fe}. The parameters are temperature dependent and were already validated by direct comparison to the ultrafast lattice dynamics in iron, nickel and cobalt \cite{ZahnNi, ZahnCo_Fe}. The electronic and lattice heat capacities agree very well with experimental data \cite{Hofmann1956,Meschter1981}. The magnetization dynamics is calculated using the M3TM for finite spin values, for details of the M3TM see Supplemental Material and Ref. \cite{Beens2019}. Within this model, the rate parameter defining the magnetization dynamics scales linearly with the so-called spin-flip probability $a_{\rm{sf}}$. 
Here, we use $S=1/2$ for nickel, $S=2$ for iron and $S=3/2$ for cobalt \cite{Koebler2003}. For these values, within the MFA, the equilibrium magnetization as function of temperature is well reproduced for all three $3d$ ferromagnets. 
\begin{figure}[h]
    \begin{center}
\includegraphics[width=0.95\linewidth]{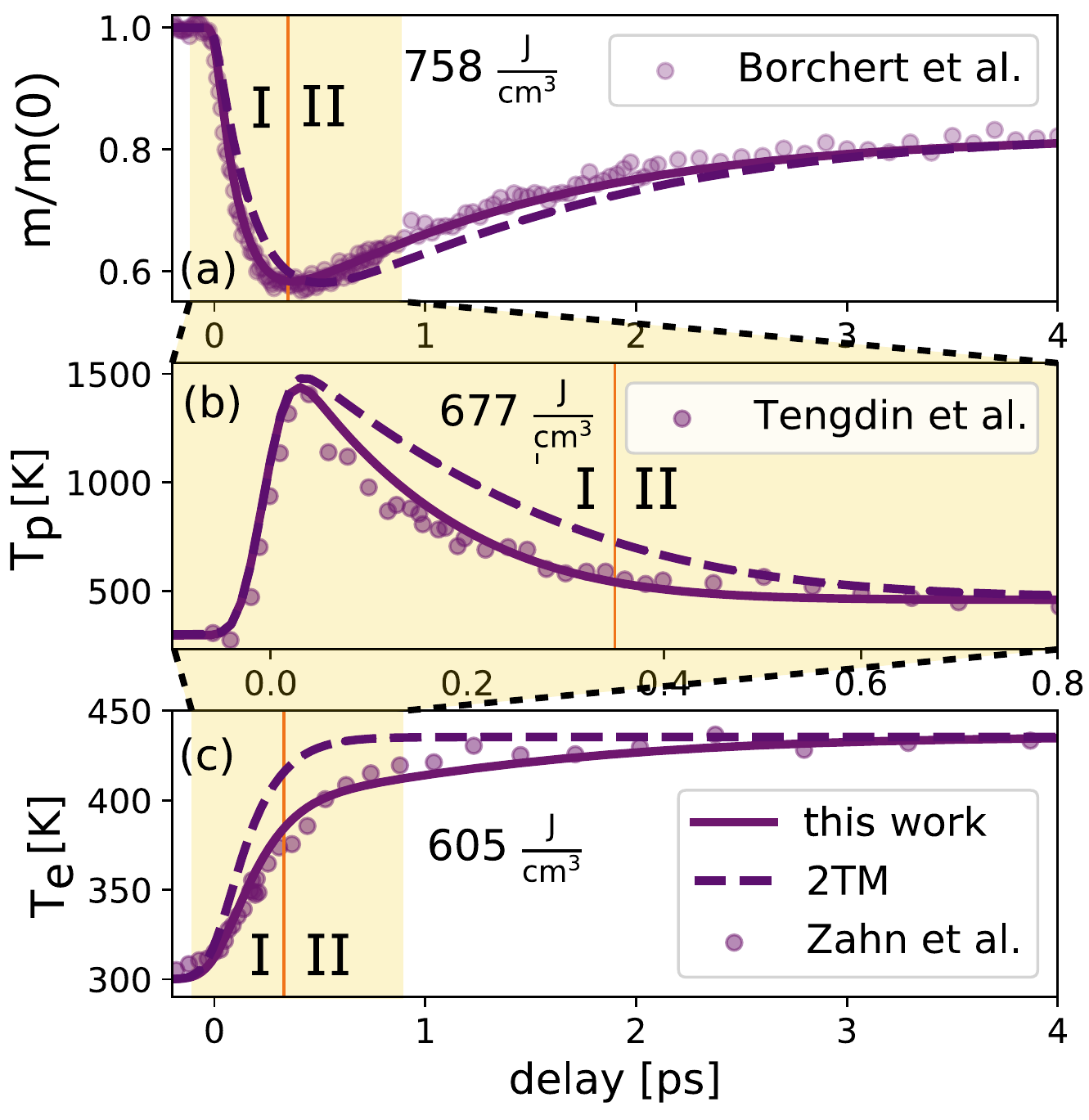}
    \end{center}
    \caption{Ultrafast dynamics in nickel; comparison between experiments (symbols) and model (solid line). For completeness, dashed line corresponds to the model with $\dot{Q}_{\rm{e-s}}=0$, absorbed energies from our simulations are indicated, the simulated pump fluences were reduced to fit simulations with $\dot{Q}_{\rm{e-s}}=0$. 
(a) Magnetization dynamics in nickel measured by MOKE \cite{Borchert2020}.
(b) Electron temperature dynamics measured by tr-ARPES \cite{Tengdin2018}.
(c) Lattice temperature dynamics measured by electron diffraction \cite{ZahnNi}. 
The orange vertical lines mark the transition between phases I and II.}
\label{fig:Fig2}
\end{figure}
Figure \ref{fig:Fig2} shows an example of the comparison of our model to experimental data for the magnetization [Fig. \ref{fig:Fig2} (a)], electron [Fig. \ref{fig:Fig2} (b)] and lattice [Fig. \ref{fig:Fig2} (c)] temperature dynamics in nickel. Vertical orange line  denotes the transition from phase I ($dm/dt<0$) to phase II ($dm/dt>0$).
The electron temperature is retrieved from tr-ARPES measurements on a $400$ nm nickel film, where the electronic energy distribution around the Fermi edge is measured and fitted to the Fermi-Dirac distribution, yielding the electronic temperature \cite{Tengdin2018}. The lattice dynamics are retrieved from time resolved electron diffraction experiments on $20$ nm thin films on a Si$_3$N$_4$ substrate. Intensity changes in the diffraction patterns reveal the mean square displacement, from which the lattice temperature can be deducted \cite{ZahnNi}. Finally, the magnetization dynamics were retrieved from time-resolved MOKE experiments on $15$ nm thin films on glass wafer substrate, where the Kerr rotation, proportional to the sample magnetization, was recovered by amplitude changes in the hysteresis loops \cite{Borchert2020}.
By direct comparison of the magnetization dynamics to the model [solid line in Fig. \ref{fig:Fig2} (b)], we obtain a value for $a_{\rm{sf}}$.
Figure \ref{fig:Fig2} (a) shows the magnetization dynamics calculated with the M3TM for $a_{\rm{sf}}=0.09$ (solid line), and experiment (symbols) \cite{Tengdin2018}.
The exact value of the absorbed fluence in experiments is difficult to estimate. We use a value in our model for the absorbed fluence such that the final temperature, and therefore the final magnetic state for one exemplary fluence, matches the experiment. 
The dashed line in Fig. \ref{fig:Fig2} shows the result of the model by assuming $\dot{Q}_{\rm{e-s}}=0$ in Eq. \eqref{eq:TTM1}. Although the effect of this term on the magnetization dynamics is marginal [Fig. \ref{fig:Fig2} (a)],  it becomes critical for the correct description of the electron and lattice temperature dynamics [Fig. \ref{fig:Fig2} (b) and (c)]. 
Symbols in Figs. \ref{fig:Fig2} (b) and (c) show the electron and lattice temperature dynamics, measured in Ref. \cite{Tengdin2018} and \cite{ZahnNi}, respectively. We note that for $\dot{Q}_{\rm{e-s}}=0$, the electron system cools down slower than experimental data while the lattice heats up faster.  This behaviour cannot be described by adjustment of $g_{\rm{ep}}$, which will accelerate the cool down of the electron temperature, but also the heat up of lattice temperature. 
The correction provided by $\dot{Q}_{\rm{e-s}} \sim \dot{m}$ strongly depends on the magnetization dynamics, which in turn is determined by $a_{\rm{sf}}$ ($\dot{m} \sim a_{\rm{sf}}$). Thus, one can regulate through $a_{\rm{sf}}$ the speed-up of the electron temperature cooling by increasing the transfer of energy from the electron to the magnetic system. This procedure has been used in previous works within the framework of atomistic spin dynamics simulations \cite{ZahnNi,ZahnCo_Fe}, however we emphasize here that in those works a direct comparison to experimentally measured magnetization dynamics in similar samples and conditions was missing.  
Here, we use the data recorded by Borchert et al. \cite{Borchert2020} for the three 3$d$ ferromagnets, for which both the samples and the conditions are practically the same as the one used to measure the lattice temperature dynamics \cite{ZahnCo_Fe}. 
First, we set the constraint of using a value for $a_{\rm{sf}}$ that reproduces the observed magnetization dynamics [Fig. \ref{fig:Fig2}(a)]. 
Second, we find that by using exactly the same set of parameters that reproduces the magnetization dynamics, the lattice and electron temperature dynamics are also very well reproduced by our model [Figs. \ref{fig:Fig2}(b,c)].
A second step to validate our model for nickel is to extend the comparison to experimental data for different fluences.  

\begin{figure}[t!]
    \centering
    \includegraphics[width=.47\textwidth]{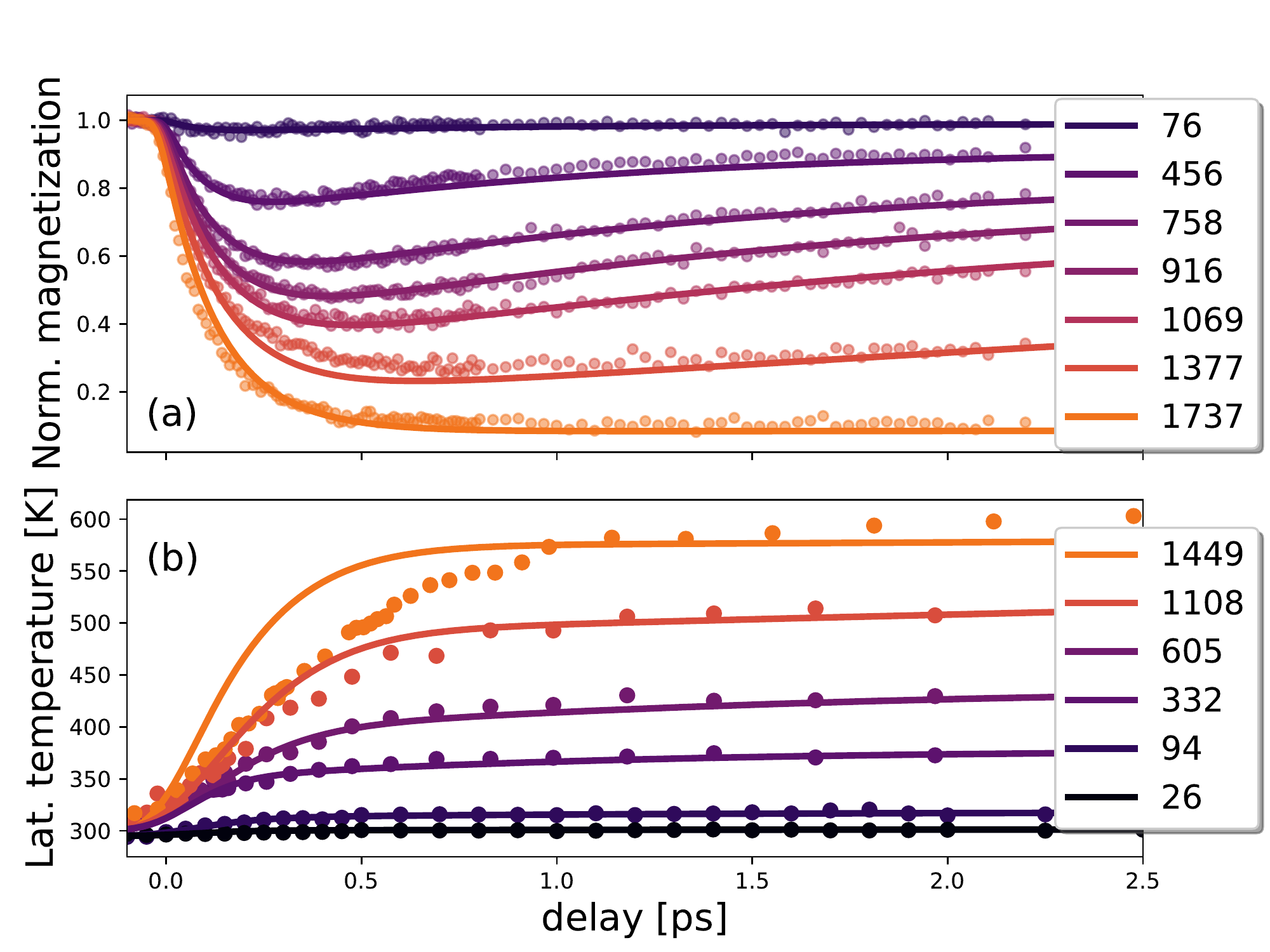}
    \caption{
    Magnetization and lattice dynamics in nickel for a range of pump fluences. 
    (a) Magnetization dynamics in nickel from experiments by Borchert et al. \cite{Borchert2020} (dots) and model (lines). The legend shows the total absorbed energy density in units of J/cm$^3$. (b) Lattice dynamics in nickel from experiments by Zahn et al. \cite{ZahnNi} (dots) and model (lines). The color scheme is matched to similar absorbed energies as in panel (a).}
    \label{fig:Fig3}
\end{figure}

Figure \ref{fig:Fig3}(a) shows magnetization dynamics, tr-MOKE measurements in symbols and model results in lines, for nickel upon irradiation with pump fluences ranging from $0.5-15 \rm{mJ}/\text{cm}^2$ \cite{Borchert2020}. As the effect of incident fluences are hardly comparable across different experimental setups, we denote the fits by the absorbed energy density used in the model calculations. 
We note that all the parameters in the model are the same as those used in Fig. \ref{fig:Fig2}. The only parameter that changes is the value of the absorbed power, $S_0$, which exactly reflects the ratio of incident fluences from experiments. 
Our model replicates the experimental magnetization data perfectly [Fig. \ref{fig:Fig3}(a)] for different incident fluences. The lattice temperature dynamics is also reproduced by our model to a high level of accuracy [Fig. \ref{fig:Fig3}(b)]. 
We find some discrepancy between the model and lattice temperature dynamics for the highest value of the absorbed energies. We argue that this comes as a result of the MFA expression for the rate of the electron-spin energy flow $\dot{Q}_{\rm{e-s}}$. 
For high fluence, the system reach elevated temperatures. This is reflected by the low values of the magnetization that are achieved. For these situations critical spin fluctuations become important, which in turn are poorly captured by the MFA. We note that this discrepancy was not observed  in a previous work using ASD simulations \cite{ZahnNi}. Atomistic spin dynamics methods account correctly for the contribution to the energy of the spin fluctuations at elevated temperature. However, ASD simulations are unable to reproduce the experimental magnetization dynamics.
By contrast, the M3TM model, with only one free parameter, $a_{\rm{sf}}$, provides an excellent agreement to experimental observations. One remaining question for nickel is the origin of the value of $a_{\rm{sf}}=0.09$. 
With a generalization of spin- and energy-dependent Eliashberg functions, Carva et al. \cite{Carva2011} have computed the spin-flip probabilities upon electron-phonon scattering of Elliott Yafet type in relativistic DFT. For nickel, they have found a range of values, $a_{\rm{sf}}=0.04-0.09$, which agrees with our model.
The transition rates are strongly dependent on the occupation-difference of electronic majority and minority states, which only arises for non-thermal electron distributions.
Our model treats the electron and lattice systems in thermal equilibrium  throughout the process. The spin system however is neither at equilibrium with the electronic system, nor internally. Thus, it describes spin-flips and magnon excitations of which the energetic cost is compensated by a shift in the close to equilibrium electron distribution. The influence of non-thermal electron distributions is an ongoing discussion \cite{Chekhov2021, Tengdin2018, Carva2011}. A final question is about the applicability of this model to the other two 3$d$ ferromagnets, iron and cobalt. 

\begin{figure}[t]
%    \centering
    \includegraphics[width=1\linewidth]{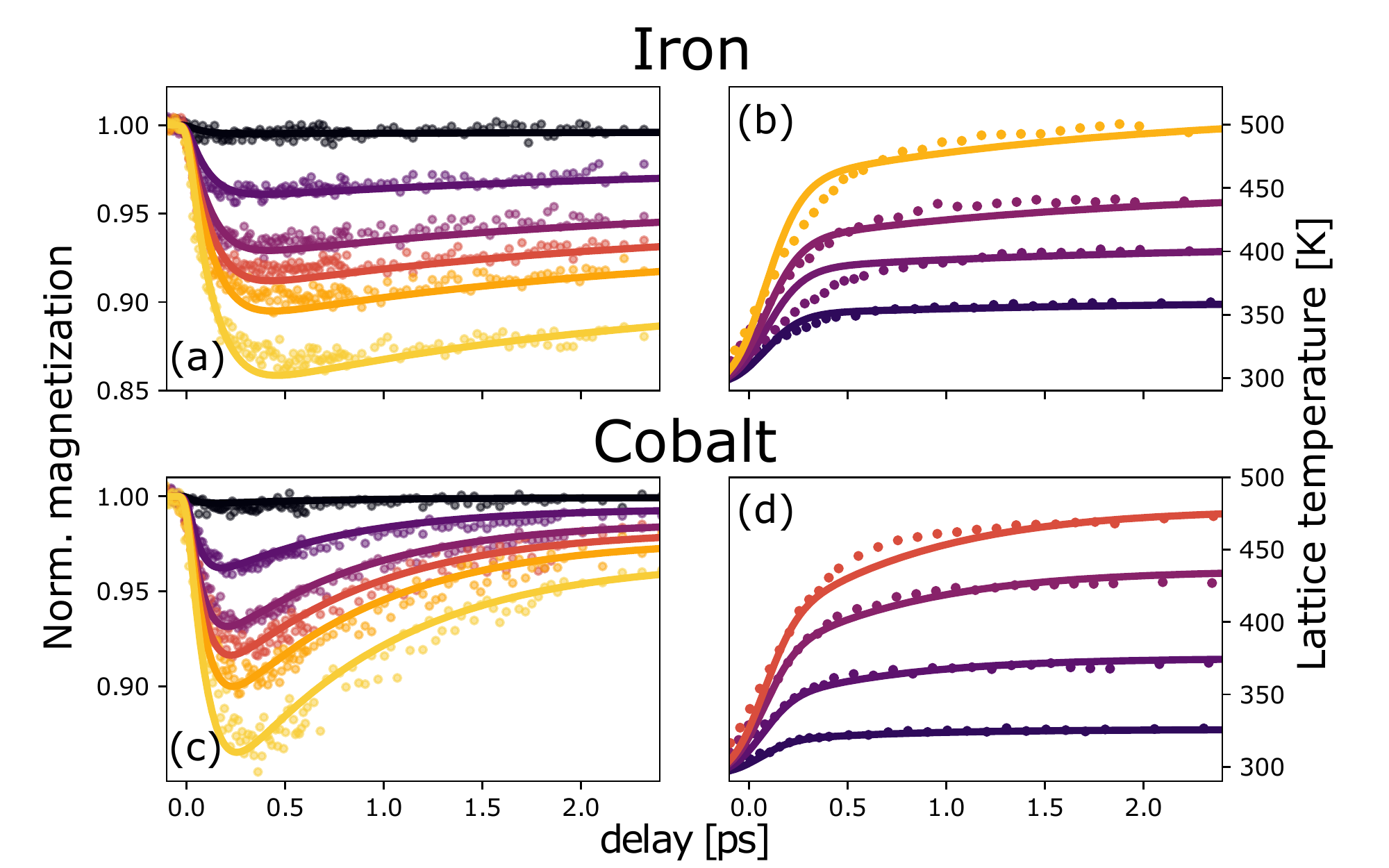}
    \caption{(a) and (b): Magnetization and lattice dynamics in cobalt. Simulations (lines) in comparison to experimental data (dots) from Borchert et al. \cite{Borchert2020} and Zahn et al. \cite{ZahnCo_Fe}, respectively. The color coding corresponds to similar absorbed pump energies, listed in methods section. (c) and (d): Magnetization and lattice dynamics in iron.}
    \label{fig:Fig4}
\end{figure}

The model describes the experimentally measured magnetization and lattice temperature dynamics in iron and cobalt. The parameters in the 2TM [Eqs. \eqref{eq:TTM1} and \eqref{eq:TTM2}] have been estimated through ab-initio calculations and validated in a previous work \cite{ZahnCo_Fe}. We estimate the  spin-flip probabilities for iron and cobalt by comparing the magnetization dynamics to the model for one single laser fluence, similar to the procedure used for nickel  [see Fig. \ref{fig:Fig2}]. 
Figure \ref{fig:Fig4} shows the comparison between the model and experiments for both the magnetization and lattice temperature dynamics in iron and cobalt. 
The color coding in  Fig. \ref{fig:Fig4} scales with the amount of absorbed energy density from the laser pulse used in Eqs. \eqref{eq:TTM1} and \eqref{eq:TTM2}.  
For iron, the magnetization dynamics is well reproduced by the model [Fig. \ref{fig:Fig4} (a)]. 
We note that in literature, experimentally measured magnetization dynamics deviate significantly from one another depending on the sample characteristic and experimental technique\cite{Carpene2008, Chekhov2021, Zahnphd}.
The lattice temperature dynamics  is reproduced with a satisfactory degree of accuracy [Fig. \ref{fig:Fig4} (b)]. 
From our model, the estimated spin-flip probability for iron $a_{\rm{sf}}=0.07$ agrees well with ab-initio calculations of $a_{\rm{sf}}=0.04-0.07$ \cite{Carva2013}.
For cobalt, the agreement between the experimentally measured magnetization dynamics and model is excellent [Fig. \ref{fig:Fig4} (a)]. The agreement to the lattice temperature dynamics is also excellent. Our estimation for the spin-flip probability $a_{\rm{sf}}=0.07$ is  larger than  the ab-initio calculations, $a_{\rm{sf}}=0.01-022$, albeit in the same order of magnitude.
Overall, our model provides a very good match for the spin-flip probability in comparison to the ab-initio calculations for the three 3$d$ ferromagnets (see Table \ref{tab:asf}), which strongly supports our claim that the electron-phonon mediated spin-flip is the main microscopic mechanism driving the ultrafast magnetization dynamics in 3$d$ ferromagnets.

%We observe that the model results for the lattice dynamics start to deviate for the higher fluence values. This could arise from the ab-initio calculations of heat capacities and electron-phonon coupling, which are calculated close to thermal equilibrium. Moreover, the set-ups used for electron diffraction experiments and tr-MOKE could also generate some differences, so that the parameters vary slightly. In this work, for the sake of completeness, we have preferred to use the same parameters for both experiments.

To summarize, we have found evidence of electron-phonon mediated spin-flips as the intrinsic driving mechanism for the ultrafast magnetization dynamics in all three 3$d$ ferromagnets, nickel, iron and cobalt. By direct comparison between our model and experimental electron and lattice temperature as well as magnetization dynamics, we determine the spin-flip probability for each material, which in turn agrees with ab-initio predictions. Our model supports ultrafast transfer of spin angular momentum from the spin to lattice system, in line with the experimental work of Tauchert et al. \cite{Tauchert2022} in nickel. They have found that the angular momentum loss during the ultrafast demagnetization is transferred to the lattice in the form of chiral phonons.

\begin{table}[t]
\caption{\label{tab:asf}%
Spin-flip probabilities $a_{\rm{sf}}$ from ab-initio calculations \cite{Carva2013}, previous M3TM fits \cite{Koopmans2010} and this work.}
\begin{ruledtabular}
\begin{tabular}{cccc}
\textrm{3$d$-ferromag.}&
\textrm{Carva et al.\cite{Carva2013} }&
\textrm{Koopmans et al. \cite{Koopmans2010}}&
\textrm{this work}\\
\colrule
Nickel & 0.04-0.09 & 0.17-0.2 & 0.09\\
Iron & 0.04-0.07 &  & 0.07\\
Cobalt & 0.01-0.022 & 0.135-0.165 & 0.07\\
\end{tabular}
\end{ruledtabular}
\end{table}

\emph{Acknowledgments.--} We gratefully acknowledges funding by the Deutsche Forschungsgemeinschaft (DFG, German Research Foundation)—Project-ID 328545488—TRR 227, Project No. A08. U. A. gratefully acknowledges support by grant PID2021-122980OB-C55 and the grant RYC-2020-030605-I funded by MCIN/AEI/10.13039/501100011033 and by “ERDF A way of making Europe” and “ESF Investing in your future”.

\bibliography{main}% Produces the bibliography via BibTeX.

\clearpage 
\widetext 

\setcounter{figure}{0}
\setcounter{page}{1}
\setcounter{equation}{0}
\renewcommand{\thepage}{S\arabic{page}} 
\renewcommand{\theequation}{S\arabic{equation}} 
\renewcommand{\thesection}{S\arabic{section}}
\renewcommand{\thesubsection}{S\arabic{subsection}}
\renewcommand{\thetable}{S\arabic{table}}  
\renewcommand{\thefigure}{S\arabic{figure}}

\textbf{Supplementary Material to "Electron-phonon mediated spin-flip as driving mechanism for ultrafast magnetization dynamics in 3$d$ ferromagnets"}

\section{S1. Microscopic three temperature model: magnetization dynamics}

The magnetization dynamics are determined by the statistical change of occupation numbers $f_{m_{s}}$ of the $S_z$ component $m_s$:

\begin{eqnarray}
\frac{dm}{dt} &=& -\frac{1}{S} \sum_{ms=-S}^{ms={+S}}m_s\frac{df_{m_s}}{dt}\\
\frac{df_{m_s}}{dt} &=& -(W^+_{m_s}+W^-_{m_s})f_{m_s}+W^+_{m_{s-1}}f_{m_{s-1}}+W^-_{m_{s+1}}f_{m_{s+1}}\\
W^{\pm}_{m_s} &=& R\frac{Jm}{4Sk_BT_c}\frac{T_p}{T_c}\frac{\text{e}^{\mp \tfrac{\tiny{Jm}}{\tiny{2Sk_BT_e}}}}{\text{sinh}(\frac{\tiny{Jm}}{\tiny{2Sk_BT_e}})} \nonumber \\
&& \cdot (S(S+1)-m_s(m_s\pm 1)),
\end{eqnarray}

The analytical description of the transition rates $W^{\pm}_{m_s}$ was introduced by Beens et al. \cite{Beens2019}. $J=3 k_B T_C S/(S+1)$ is the mean flied exchange parameter for a system of spin $S$ and $T_C$ the Curie temperature. The rate parameter

\begin{align}
    R= \frac{8a_{\rm{sf}}{g_{\rm{ep}}} k_{B} T^2_C V_{\rm{at}}}{(\mu_{\rm{at}}{E^2_{\rm{D}}})}
\end{align} 
is proportional to the microscopic spin-flip probability $a_{\rm{sf}}$, where $V_{\rm{at}}$ is the atomic volume in the lattice and $E_D=k_B T_D$ the Debye energy.

\end{document}